\documentclass[twocolumn,english,aps,reprint, superscriptaddress,showpacs,showkeys]{revtex4-2}
\usepackage[utf8]{inputenc}
\usepackage{amsmath,amssymb,bbm,mathrsfs,bm,braket,color,graphicx,comment,xcolor}
\usepackage[colorlinks,citecolor=blue,urlcolor=blue,linkcolor = blue]{hyperref}
\usepackage[caption=false]{subfig}

\newcommand{\bfq}{{\boldsymbol{q}}}
\newcommand{\bfbeta}{{\boldsymbol{\beta}}}
\newcommand{\bfgamma}{\boldsymbol{\gamma}}
\newcommand{\bfOmega}{\boldsymbol{\Omega}}
\newcommand{\calC}{\mathcal{C}}

\newcommand{\squaredots}{
    \vspace{-.175em}
    \tikz[line cap=round, line join=round]{
    \draw[black] (0ex,0ex) -- (0ex,0.8ex) --  (0.8ex,0.8ex) --  (0.8ex,0ex) -- cycle;
    \draw[color=black, fill=black] (0ex,0ex) circle (0.175ex);
    \draw[color=black, fill=black] (0ex,0.8ex) circle (0.175ex);
    \draw[color=black, fill=black] (0.8ex,0ex) circle (0.175ex);
    \draw[color=black, fill=black] (0.8ex,0.8ex) circle (0.175ex);
    }
}

\usepackage{tikz}

\begin{document}
\title{Error Mitigation for Quantum Approximate Optimization}

\author{Anita Weidinger}
\affiliation{Institute for Theoretical Physics, University of Innsbruck, A-6020 Innsbruck, Austria}

\author{Glen Bigan Mbeng}
\affiliation{Institute for Theoretical Physics, University of Innsbruck, A-6020 Innsbruck, Austria}

\author{Wolfgang Lechner}
\email{wolfgang@parityqc.com\\wolfgang.lechner@uibk.ac.at}
\affiliation{Institute for Theoretical Physics, University of Innsbruck, A-6020 Innsbruck, Austria}
\affiliation{Parity Quantum Computing GmbH, A-6020 Innsbruck, Austria}

\begin{abstract}
Solving optimization problems on  near term quantum devices requires developing error mitigation techniques to cope with hardware decoherence and dephasing processes. We propose a mitigation technique based on the LHZ architecture. This architecture uses a redundant encoding of logical variables to solve optimization problems on fully programmable planar quantum chips.  We discuss how this redundancy can be exploited to mitigate errors in quantum optimization algorithms. In the specific context of the quantum approximate optimization algorithm (QAOA), we show that errors can be significantly mitigated by appropriately modifying the objective cost function.
\end{abstract}

\maketitle

\section{Introduction}
In recent years immense effort has been made to leverage quantum computers to solve industry relevant optimization problems \cite{ Albash_Lidar_RevModPhys2018_ReviewAQC, Zhou_2020_PhysRevX_QAOAPerformance, Moll_QuSciAndTechnology2018_VQAandOpt, Guerreschi_ScientificReports2019, Harrigan_2021_NaturePhysics}. 
The main obstacles to observing a quantum advantage are the limitations of the current noisy intermediate-scale quantum (NISQ) devices, which offer only a few hundred qubits and are prone to errors and noise \cite{Preskill_2018_Quantum_NISQ, Stilck_2021_NaturePhysics_LimitationsNISQ}. Until technology advances, specific algorithms have been designed to utilize the maximum out of the imperfect hardware \cite{Bharti_2022_RevModPhysics_NISQAlgorithms}. The proposed algorithms generally follow a hybrid quantum-classical approach, simultaneously exploiting quantum and classical computational power. The most prominent hybrid algorithms in current research are the variational quantum algorithms (VQAs), where parameterized quantum circuits are combined with a classical optimizer \cite{Cerezo_2021_NatureReviewPhysics_ReviewVQA, McClean_2016_NewJournalofPhysics_VQA}. VQAs are suitable for various applications like chemistry \cite{Kandala_Nature2017_VQAChemistry, Arute_Science2020_VQAChemistry} or machine learning \cite{Biamonte_Nature2017_QuantumMachineLearning, Mitarai_PhysRevA2018_VQAMachineLearning}. In particular, the quantum approximate optimization algorithm (QAOA) is a promising VQA that aims at finding approximate solutions to combinatorial optimization problems \cite{Farhi_2014_arXiv, Zhou_2020_PhysRevX_QAOAPerformance}. 
Hybrid algorithms have the advantage of providing shallow circuit depths, which makes them less vulnerable to noise. Gates are potential error sources, however, in an ideal circuit, more circuit layers would improve the quality of the solution. In practice, there is a break-even point, where adding a layer negatively affects the algorithms' performance. First experimental setups showed how severely limited the circuit depth, and therefore the algorithm's performance, is in current devices \cite{Harrigan_2021_NaturePhysics, Pagano_2020_trappedIonSimulator, Lacroix_2020_PRX_ContinousGateSet}. 
Hence, it is essential to lessen the effect of noise to exploit the capabilities of VQAs fully. Various quantum error mitigation (QEM) techniques have been developed to tackle this issue. In general, QEM aims to reduce the impact of noise via classical post-processing and multiple circuit runs, requiring no qubit overhead. \cite{Endo_2021_JournalPhysSocietyJapan_ReviewHybridAlgQEM, Cai_OBrien_arxiv2022_ReviewQEM}. These resource sparing techniques are suitable for NISQ devices, in contrast to quantum error correction (QEC) codes, which instead require significant qubit overheads \cite{Gottesman_ApplMathematics2010_QEC, 
Raussendorf_Harrington_PhysRevLett2007_QEC}.
Some examples of QEM schemes are 
extrapolation and quasi-probability methods \cite{Li_2017_PHysRevX_extrapolation, Temme_2017_PhysRevLetters_ExtrapolationQuasiProb, Endo_2018_PhysRevX_EMnearFutureApplications},  quantum subspace expansion \cite{McClean_2017_PhysRevA_SubspaceExpansion, McClean_2020_NatureCommunications_SubspaceExpansions}, individual error reduction \cite{Otten_2019_npjQI_IndividualErrorReduction}, symmetry verification \cite{BonetMonroig_2018_PhysRevA_SymmetryVerification, Czarnik_2021_ARXiv_SymmetryVerification, McArdle_2019_PhysRevLetters_EMDigitalQuantumSimulations, Botelho_2022_PhysRevA_EMMidCircMeasurment} and combinations like error extrapolation with symmetry verification \cite{McArdle_2019_PhysRevLetters_EMDigitalQuantumSimulations, Cai_2021_npjQI_QEMCombined} and quasi-probability method \cite{Cai_2021_npjQI_QEMCombined}.

In this work we propose a novel error mitigation technique for QAOA that is based on the LHZ- or parity architecture  \cite{Lechner_2015_ScienceAdvances}. Instead of using the energy as cost function of the variational algorithm, our method introduces logical qubits and the decoded \textit{logical energy} is used. This \textit{logical energy} is the result of the evaluation of spanning trees in the parity variables, where each physical qubit contributes to multiple logical qubits. It is this redundancy that  introduces the error mitigation features of the method. 

The parity architecture was initially designed to tackle the issue of limited connectivity in quantum annealing hardware. However, when combined with digital hardware, the parity architecture provides the benefit of full parallelization of quantum gates \cite{Lechner_2020_IEEE}, and universal quantum computing \cite{Fellner_Lechner_PhysRevLett2022_UniversalQC}.
In addition, Ref.~\cite{Pastawski_2016_PhysRevA_ECandLHZ} shows that the parity architecture can be viewed as a classical low-density parity check code, and adding classical post-processing procedures like belief propagation make it more robust against noise. 
Further, Ref.~\cite{Rocchetto_2016_ScienceAdvances_LHZStabilizer} establishes a stabilizer-based formalism for the parity architecture. 

The transformation from logical to physical qubits introduces redundant information which can be exploited in classical post-processing to correct errors. Our method uses this redundancy not just in post-processing but also during computation. Consequently, the noise-reducing benefit is propagated throughout the algorithm and not just instilled in the end.
This is done by redesigning the information that is handed from the quantum circuit to the classical optimizer, represented by the cost function. With this adaption, the computation is performed on the physical quantum hardware, while the classical optimizer always stays in the logical subspace. Our results demonstrate that this modification introduces quantitative error mitigation for QAOA in the presence of noise.
The next sections are organized as follows: In Sec.~\ref{sec:methods} we outline the theoretical background for this paper, which includes a description of QAOA, parity QAOA, and a detailed explanation of our novel method: decoded parity QAOA. In Sec.~\ref{sec:results}  we discuss the numerical simulations' results where we compare the discussed QAOA approaches. Finally, we discuss the open questions and the prospects of this topic.

\section{Methods}
\label{sec:methods}

\subsection{QAOA with rerouting}\label{sec:QAOA}
The Quantum Approximate Optimization Algorithm (QAOA) \cite{Farhi_2014_arXiv} aims to find approximate solutions to combinatorial optimization problems, cast in the form of energy minimization of a general $N$-spin problem Hamiltonian
\begin{align}
\begin{split}
    H_p &=\sum_{i}J_{i} \sigma_i^z + \sum_{i<j}J_{ij} \sigma_i^z \sigma_j^z\\  &+ \sum_{i<j<k}J_{ijk} \sigma_i^z \sigma_j^z\sigma_k^z + \dots, 
\end{split} \label{eq:Hp-generic}
\end{align}
where $\sigma^{\lbrace x,y,z\rbrace}_j$ denote the Pauli spin operators and $\lbrace J_i, J_{ij}, J_{ijk},\dots \rbrace$ are long-range, multi-spin interactions.

The simplest $p$-level QAOA~\cite{Farhi_2014_arXiv,Hadfield_2019_Algorithms_QAOA+} starts from an initial state $\ket{\Psi_0}=\ket{+}^{\otimes N}$ and alternates a phase separation gate $U_p(\gamma) = e^{-i\gamma\hat{H}_p}$ and a mixing gate $U_x(\beta) = \prod_{j=1}^{N} e^{-i\beta\sigma^x_j}$ for $p$ rounds in a quantum circuit. Running the circuit on suitable quantum hardware generates the variational state
\begin{equation}
    \label{eq:QAOA_circuit}
    \ket{\Psi(\bm{\beta},\bm{\gamma})} =  U_x(\beta_p) U_p(\gamma_p)\dots U_x(\beta_1) U_p(\gamma_1)\ket{\Psi_0},
\end{equation}
which depends on the $2p$ parameters $\bm{\gamma}=\{\gamma_1,\dots, \gamma_p\}$ and $\bm{\beta}=\{\beta_1,\dots, \beta_p\}$.
Then, by repeated measurements of the state in the computational basis, we can estimate the QAOA objective function
\begin{align}
\label{eq:E_GM}
    \calC(\bfbeta,\bfgamma)
    &= \bra{\Psi(\bfbeta,\bfgamma)} H_p \ket{\Psi(\bfbeta,\bfgamma)},
\end{align}
for any parameter choice $(\bfbeta,\bfgamma)$. This objective function 
coincides with the expectation value of the original classical spin glass energy on the measured $N$-bitstrings. We use a classical computer to implement a feedback loop optimization algorithm to find the optimal parameters $(\bfbeta^*,\bfgamma^*)$ that minimize $\calC(\bfbeta,\bfgamma)$. Finally, we run the quantum circuit with parameters $(\bfbeta^*,\bfgamma^*)$ and generate bitstrings providing approximate solutions of the classical optimization problem. Due to the nature of the variational ansatz, the quality of the approximate solution generated by ideal noiseless circuits increases monotonically with the depth $p$.
In particular, as discussed in Refs.~\cite{Farhi_2014_arXiv,Mbeng_Santoro_arXiv2019}, the adiabatic theorem~\cite{Messiah_Book2014} ensures that for $p\to\infty$ the algorithm converges to an exact solution of the classical problem. 

Although the mixing gate $U_x(\beta)$ is straightforward to implement with single-qubit operations, hardware with local connectivity graphs require an additional compilation step \cite{Hirata_QuantumInfoConf2011_LNN, Bako_Zimboras_arXiv2022_FUNCQAOA} to decompose the phase gate $U_p(\gamma)$ into a sequence of available local gates. In most QAOA implementations, this compilation step relies on a rerouting strategy~\cite{tket,Harrigan_2021_NaturePhysics, Mahabubul_Swaroop_2020, Lao_Almudever_IEEE2022, Li_Xie_Proceedings2019, Murali_Martonosi_Proceedings2019, Tan_Cong_Proceedings2020}, which performs additional layers of SWAP gates such that the interacting spins correspond to an edge in the hardware graph at least once, enabling $U_p(\gamma)$'s implementation. However, the rerouting introduces a SWAP-gate overhead that should be minimized~\cite{tket} to reduce the effect of decoherence and dephasing processes on the compiled quantum circuit. 

\subsection{Parity QAOA \label{sec:LHZ_QAOA}}
In this section we describe \textit{parity QAOA}~\cite{Lechner_2015_ScienceAdvances,EnderTerHoeven_2021_arXiv_Compiler,Ender_2022_PRX_ModularParity,Dlaska_Lechner_PRL2022,Rocchetto_2016_ScienceAdvances_LHZStabilizer}, which uses an alternative compilation strategy, suitable for state-of-the-art quantum devices with planar chips~\cite{Arute_2019_Nature, Henriet_2020_Quantum, Saffman_2010_RevModPhysics, Bloch_2008_RevModPhysics}. To simplify the presentation, we follow Ref.~\cite{Lechner_2015_ScienceAdvances} and specifically consider quadratic unconstrained binary optimization (QUBO)~\cite{Lucas_FrontPhys2014} problem Hamiltonians:
\begin{equation}
\label{eq:Hp-qubo}
H_p = \sum_{i=1}^{N}\sum_{j<i}J_{ij}\sigma_i^z\sigma_j^z.
\end{equation}
However, we refer to Refs.~\cite{Rocchetto_2016_ScienceAdvances_LHZStabilizer,EnderTerHoeven_2021_arXiv_Compiler} for methods to tackle the more general problem Hamiltonian of Eq.~\eqref{eq:Hp-generic} with  parity QAOA.

\begin{figure}
\includegraphics[width=\columnwidth]{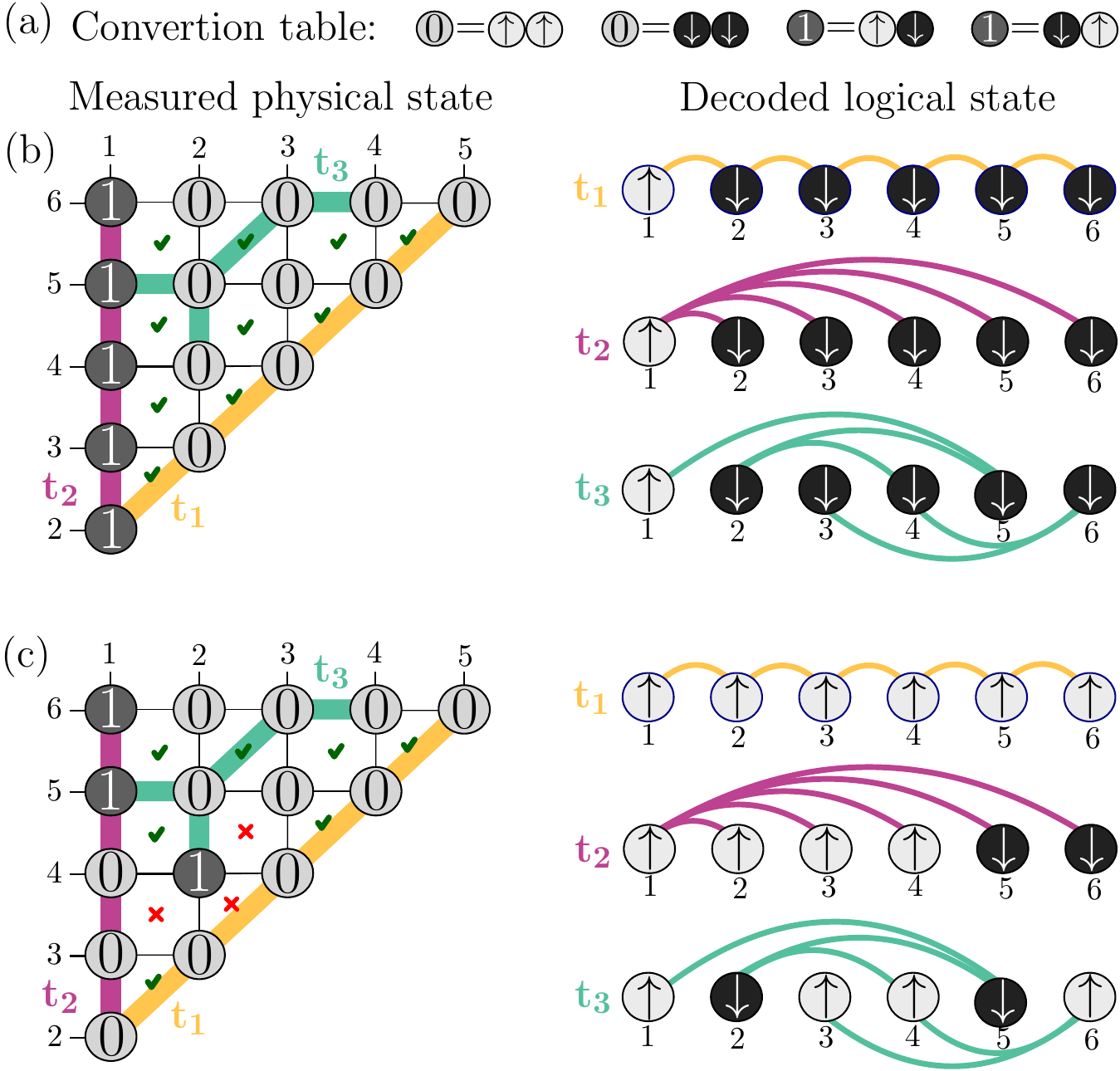}
\caption{3 Decoding examples, based on the convertion table depicted in panel (a), for given physical configurations. Panel (b) shows a constraint fulfilling physical state (left) while the one in (c)  violates some constraints (red crosses). The physical qubits used for decoding are marked with colored lines, labeled as $T_1$, $T_2$ and $T_3$.
Those 3 decoding possibilities return 3 identical logical configurations (right) for the constraint fulfilling configuration in (a) and 3 different logical states for the constraint violating configuration in (c). The used physical qubits are marked as interactions in the logical graph. They are spanning trees in the logical graph, covering all logical qubits without forming a cycle.}\label{fig:Decoding}
\end{figure}

Parity QAOA relies on the parity (or LHZ) transformation~\cite{Lechner_2015_ScienceAdvances,EnderTerHoeven_2021_arXiv_Compiler} to first encode the optimization problem into the local fields of a 2D local problem Hamiltonian. In particular, for all-to-all connected QUBO problems, the parity mapping replaces the original $N$ `logical' qubits $\sigma_i^z$ with $K=\frac{N(N-1)}{2}$ physical parity qubits $\Tilde{\sigma}_{\nu}^z$ each representing the relative configuration of two logical qubits $\sigma_i^z\sigma_j^z\to \Tilde{\sigma}_{i+(N-1)j}^z$ [see Fig.~\ref{fig:Decoding}(a)]. In terms of the new physical qubits, the problem Hamiltonian of Eq.~\eqref{eq:Hp-qubo} takes the simple local field form
\begin{align}
    \label{eq:HZ}
    H_p\to \tilde{H}_{p} =  \sum_{\nu=1}^{K} \tilde{J}_{\nu}\Tilde{\sigma}_{\nu}^z,
\end{align}
with $\tilde{J}_{i+(N-1)j}=J_{ij}$. However, the parity transformation also enlarges the configuration space by introducing spurious (or invalid) 
 physical states which do not represent any logical state.  
As suggested in Ref.~\cite{Lechner_2015_ScienceAdvances}, we can address this issue by arranging the qubits on a planar grid and considering a set of $L=\frac{(N-1)(N-2)}{2}$ independent local four (and three)  body plaquette constraints terms $H_{\squaredots,l}=\Tilde{\sigma}_{(l,1)}^z\Tilde{\sigma}_{(l,2)}^z\Tilde{\sigma}_{(l,3)}^z\Tilde{\sigma}_{(l,4)}^z$ (and $H_{\squaredots,l}=\Tilde{\sigma}_{(l,1)}^z\Tilde{\sigma}_{(l,2)}^z\Tilde{\sigma}_{(l,3)}^z$). 
This construction effectively recasts the problem of minimizing the non-local $N$-spin  Hamiltonian $H_p$ into the problem of minimizing  the $K$-spin local Hamiltonian $\tilde{H}_p$, with the constraints  $\braket{\psi|\hat{H}_{\squaredots,l}|\psi}=1$ for $l=1,\dots, L$.  Figure~\ref{fig:Decoding}(b) illustrates the resulting parity architecture for a system of $K=15$ parity qubits (or $N=6$ logical qubits). It depicts an example of a valid physical configuration that fulfills all constraints (on the left), and the corresponding configuration of the logical qubits (on the right).

The simplest $p$-level parity QAOA approximates the solution of the constraint $K$-spin optimization problem by implementing the following variational state~\cite{Lechner_2020_IEEE}
\begin{equation}
\begin{split}
\ket{\tilde{\Psi}(\bfbeta,\bfgamma,\bfOmega)} &=  \tilde{U}_x(\beta_p) \tilde{U}_c(\Omega_p) \tilde{U}_z(\gamma_p)\dots\\
&\hspace{1.2cm}\dots\tilde{U}_x(\beta_1) \tilde{U}_c(\Omega_1) \tilde{U}_z(\gamma_1)\ket{\tilde{\Psi}_0},
\end{split}
\label{eq:Parity_circuit}
\end{equation}
where $\ket{\tilde{\Psi}_0}=\ket{+}^{\otimes K}$ is the initial state,  $\tilde{U}_x(\gamma) = e^{-i\gamma\tilde{H}_p}$ and $\tilde{U}_c(\Omega) = \prod_{l=1}^{L} e^{-i\Omega H_{\squaredots,l}}$ are the phase separation gates, and  $\tilde{U}_x(\beta) = \prod_{\nu=1}^{K} e^{-i\beta\Tilde{\sigma}_j^x}$ is the mixing gate of the QAOA quantum circuit. 
Then, the optimal values of the $3p$ parameters $\bm{\gamma}=(\gamma_1,\dots, \gamma_p)$, $\bm{\beta}=(\beta_1,\dots, \beta_p)$ and $\bm{\Omega}=(\Omega_1,\dots, \Omega_p)$ are found using repeated measurements of the state in the computational basis to estimate and optimize the objective function 
\begin{align}
\label{eq:E_LHZ}
\begin{split}
    \tilde{\calC}(\bfbeta,\bfgamma,\bfOmega)
    &= \bra{\tilde{\Psi}(\bfbeta,\bfgamma,\bfOmega)} \tilde{H}_p \\
    &\hspace{0.5cm}+ c\sum_{l}(1-H_{\squaredots,l})\ket{\tilde{\Psi}(\bfbeta,\bfgamma,\bfOmega)},
    \end{split}
\end{align}
where the penalty strength $c$ is a positive constant  introduced to penalize invalid states and should be larger than $H_p$'s lowest energy gap~\cite{Lanthaler_Lechner_NewJPhys2021}. The objective function in Eq.~\eqref{eq:E_LHZ}
coincides with the weighted sum of the energy expectation values and the number of violated constraints of the measured $K$-bitstrings. As in most QAOA implementations, the adiabatic theorem ensures that for $p\to\infty$, the parity QAOA variational state in Eq.~\eqref{eq:Parity_circuit} can represent the exact solution of the constrained $K$-spin optimization problem~\cite{Lechner_2015_ScienceAdvances,Lechner_2020_IEEE}. 

The main advantage of parity mapping lies within the structure of the required gates. On the one hand, the gates $\tilde{U}_x(\beta)$ and $\tilde{U}_p(\gamma)$ involve only single-qubit operations.
On the other hand, the multi-qubit phase gate $\tilde{U}_c(\Omega)$ can be conveniently implemented on nearest neighbor planar chips by a constant-depth  sequence  of  parallel  controlled-NOT (CNOT) gates and single-qubit rotations~\cite{Lechner_2020_IEEE}, see also Appendix~\ref{app:sec:embedding}, or via optimized fast four-qubit gate operations~\cite{Dlaska_Lechner_PRL2022}. 
The numerical benchmarks in Ref.~\cite{Fellner_2021_arXiv} confirmed that these parity QAOA implementations require significantly fewer multi-qubit gates than QAOA with rerouting when running large problem instances on 
planar chips. 
The use of less error-prone gates favors parity QAOA when running the two algorithms on NISQ hardware. However, the lower number of qubits $N<K$ instead favors the rerouting strategy for QAOA. Although the tradeoff between the number of qubits and the number of multi-qubit gates still needs to be systematically analyzed, parity QAOA is a general alternative to the rerouting  strategy for QAOA  ~\cite{Fellner_MasterThesis2020,Ender_2022_PRX_ModularParity}. In the next section, we introduce a new decoded parity QAOA. The protocol is based on the evaluation of the energy of the decoded logical qubits instead of the actual energy of the physical qubits. 

\subsection{Decoded parity QAOA \label{sec:log_QAOA}}
In this section, we describe a decoding strategy 
to mitigate the errors arising in parity QAOA.

Although parity QAOA's target state  fulfills all parity constraints, running the algorithm on noisy quantum devices or with small $p$ generates errors associated with constraint violations in the readouts. We can partially correct the leakage errors by implementing various decoding strategies, which recover  logical states but do not guarantee the recovery of the target state. More specifically, a decoding strategy is a rule that assigns values to the $N$ logical qubits $\bfq=(q_1,\dots,q_N)$  given a readout of the $K$ physical qubits $\tilde{\bfq}=(\tilde{q}_1,\dots,\tilde{q}_K)$.  Here, following Refs.~\cite{Lechner_2015_ScienceAdvances,Albash_Lidar_PRA2016}, we consider a simple strategy based on spanning trees (subgraphs where all pairs of nodes are connected by exactly one path) defined on the logical qubits. We use subsets of physical qubits, describing spanning trees $t$
on the original logical qubits, to fully determine a configuration of the logical qubits $\bfq^{(t)}$ (up to a global spin flip).
 Then, if $\boldsymbol{\tilde{q}}$ satisfies all $L$ parity constraints, all the $N^{N-2}$ spanning trees return the same encoded logical qubits configuration [see Fig.~\ref{fig:Decoding}(a)]. On the other hand, when $\tilde{\bfq}$ does not satisfy all parity constraints,
 different spanning trees $t,t'$ may result in a different logical configuration $\bfq^{(t)}\neq\bfq^{(t'
 )}$ [see Fig.~\ref{fig:Decoding}(b)]. If the leakage error is sufficiently small, a majority vote over different spanning tree decoding strategies can correct the leakage errors that occurred during computation~\cite{Lechner_2015_ScienceAdvances}.  Ref.~\cite{Albash_Lidar_PRA2016} used this method to mitigate errors in adiabatic quantum optimization by adding a single decoding step at the end of the adiabatic protocol. We employ similar a decoding strategy to enhance parity QAOA's performace. 

To apply the spanning tree decoding strategy to  parity QAOA, we consider a set of $M$ spanning trees $T = \{t_1,t_2,\dots\,t_M\}$ and denote by  $D_{t_m}$ the associated linear decoding maps 
$ \ket{D_{t_m}\tilde{\bfq}} =  \ket{\bfq^{(t_m)}}$.
 We first implement the same parity QAOA variational state  of Eq.~\eqref{eq:Parity_circuit}. However to find the optimal variational parameters $(\bfbeta^*,\bfgamma^*,\bfOmega^*)$, we instead 
 minimize the expectation value of the decoded states' average energy:
\begin{align}
\label{eq:E_LLHZ}
\begin{split}
     \calC^{(T)}(\bfbeta,\bfgamma, \boldsymbol{\Omega})  
     &=\\
     \frac{1}{M} \sum_{m=1}^{M} &\bra{D_{t_m}\tilde{\Psi}(\bfbeta,\bfgamma,\bfOmega)} H_P  \ket{D_{t_m}\tilde{\Psi}(\bfbeta,\bfgamma,\bfOmega)},
\end{split}
\end{align}
 which we estimate by repeatedly measuring the parity QAOA circuit and decoding the outcome. Finally, to generate approximate solutions, we prepare the optimal variational state $\ket{ \tilde{\Psi}(\bfbeta^*,\bfgamma^*,\bfOmega^*)}$, we measure and decode the qubits $\boldsymbol{\tilde{q}}\to\{\boldsymbol{q}^{(t_1)}, \boldsymbol{q}^{(t_2)}, ..., \boldsymbol{q}^{(t_M)}\}$, and we return the best decoded logical state 
 \begin{equation}\label{eq:outputted_decoded_string}
    \bfq^{(T)}=\underset{t \in T}{\arg\min}\braket{\bfq^{(t)}|H_p|\bfq^{(t)}}.    
 \end{equation}
The output distribution is described by the quantum state $\ket{D_{T}\tilde{\Psi}(\bfbeta^*,\bfgamma^*,\bfOmega^*)}$, where $D_{T}$ is the linear decoding map 
$ \ket{D_{T}\tilde{\bfq}} =  \ket{\bfq^{(T)}}$.
The objective function Eq.~\eqref{eq:E_LLHZ} and the decoding map $D_{T}$ Eq.~\eqref{eq:outputted_decoded_string} play an important role in decoded parity QAOA. However, their definition is not unique~\cite{Barkoutsos_2020_Quantum_ObjectiveFct,Kolotouros_2022_PhysRevResearch_ObjectiveFct}. In Appendix~\ref{sec_app:different_strategies}, we study the effect of choosing a different decoding map $D_{T}$.

\section{Results and Simulations}
\label{sec:results}
In the following sections, we present numerical benchmarks of the decoded parity QAOA (Sec.~\ref{sec:log_QAOA}) and compare it to the rerouting method (Sec. \ref{sec:QAOA}). 

We benchmark QAOA on random QUBO problem instances, described by the Hamiltonian in Eq.~\eqref{eq:Hp-qubo} with uniformly distributed couplings $J_{ij}\in\{\pm 0.1, \pm 0.2, ..., \pm 1.0\}$. In particular, we consider problems with $N=3, 4, 5,6,7$ logical qubits for the rerouting method, which correspond to $K=3, 6, 10,15,21$ physical qubits for the parity architecture.

To evaluate the algorithm's performance, we first find the target ground state $\bfq^{(\mathrm{gs})}$ configuration by brute force. Then, for QAOA, we generate 100 random start parameters $\bfbeta$ and $\bfgamma$ for the rerouting strategy and $\bfbeta, \bfgamma$ and $\bfOmega$ for the parity strategy, followed by parameter updates using the Metropolis method. The run with the lowest energy $\calC(\bfbeta^*,\bfgamma^*)$ or $\Tilde{\calC}(\bfbeta^*,\bfgamma^*,\bfOmega^*)$ is used for later calculations.  We use the qiskit library~\cite{Qiskit} to simulate the quantum circuits and estimate the probability
of outputting the ground state
\begin{align}\label{eq:ground_state_probability}
    P_{\mathrm{gs}} &= |\braket{\bfq^{(gs)}|\Psi_{\mathrm{out}}}|^2,
\end{align}
where the state $\ket{\Psi_{\mathrm{out}}}$ describes the algorithm's output distribution. 
Specifically, we have  ${\ket{\Psi_{\mathrm{out}}}=\ket{\Psi(\bfbeta^*,\bfgamma^*)}}$ for QAOA with rerouting and ${\ket{\Psi_{\mathrm{out}}}=\ket{D_{T}\tilde{\Psi}(\bfbeta^*,\bfgamma^*,\bfOmega^*)}}$ for the decoded parity QAOA. 

Since the parity strategy requires more qubits than rerouting, we cannot use  $P_{\mathrm{gs}}$ to directly compare the algorithms. To enable a fairer comparison, we run the algorithms on multiple copies in parallel and take the best outcome. Assuming the same fixed budget of $K$ physical qubits for both algorithms, the resulting success probability is
\begin{equation}
\label{eq:mod_PGS}
    P_{S} = 1 - (1-P_{gs})^r,
\end{equation}
where the number of copies is $r=1$ for parity strategy or $r=K/N=\frac{N-1}{2}$ for the rerouting strategy. 
For further details on the numerical simulations, we refer to Appendix~\ref{sec_app:simulation_details}. 

\subsection{Noiseless simulations}
\label{sec:noiseless}
First, we investigate how the performance of parity QAOA depends on the number of spanning trees $M$ used. Then we fix the number of spanning trees $M$ and simulate different system sizes $N$. We compare the later results to the rerouting strategy.

\begin{figure}
 \includegraphics[width=\columnwidth] {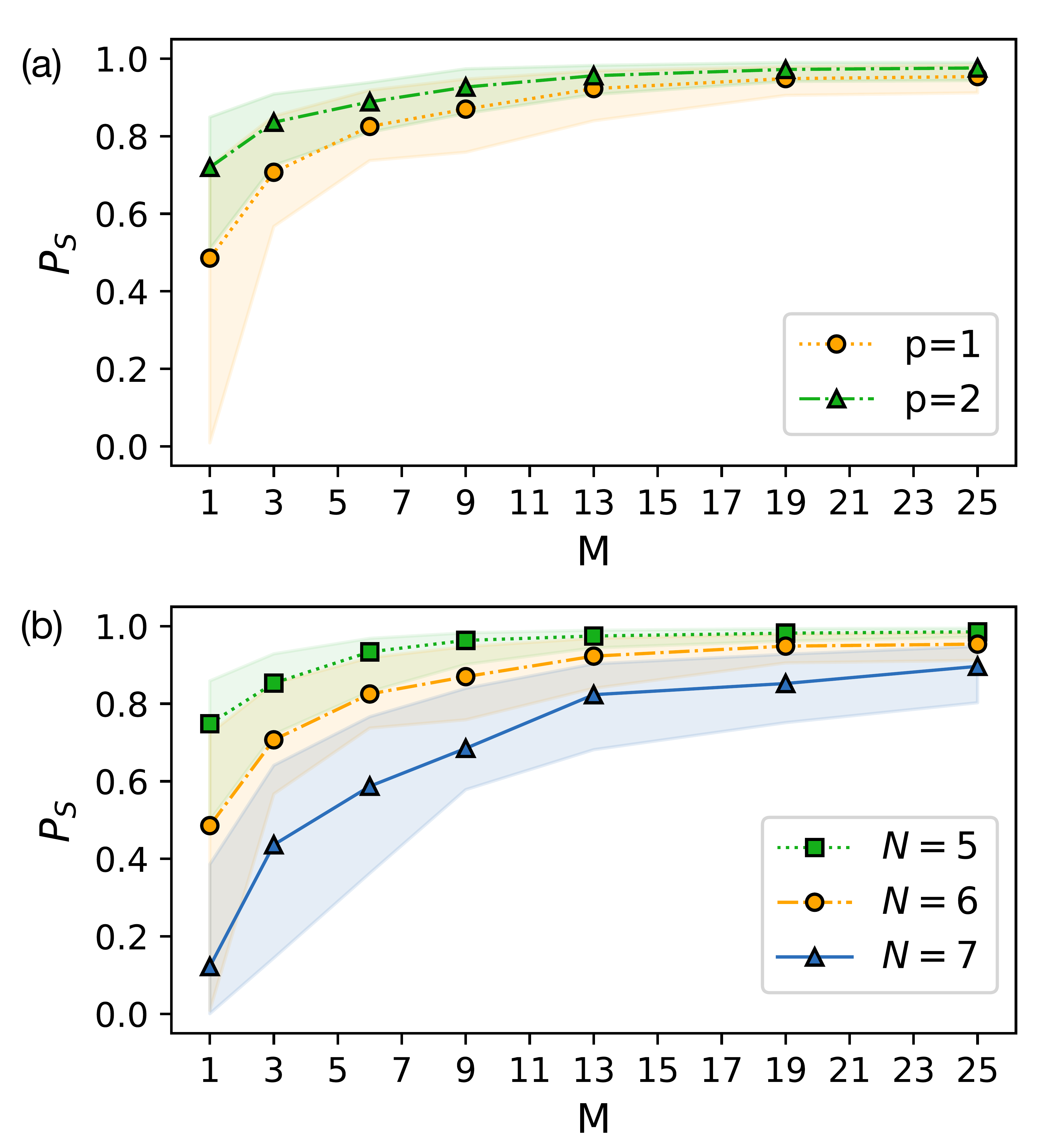}
\caption{Success probability $P_S$ for parity QAOA in dependence of the total number of spanning trees $M$ used for decoding for (a) $N=6$,  $p=1,2$ and (b) different system sizes $N$ with $p=1$. The (dashed) lines and markers represent the median, the shaded area the 25th and 75th percentile of 100 random instances.}
\label{fig:SpanningTrees}
\end{figure}

\subsubsection{decoded parity QAOA}
\label{sec:Spanning_trees}

We randomly create a set of spanning trees ${T = \{t_1,t_2,\dots\,t_M\}}$, for different $M$. More details can be found in Appendix~\ref{sec_app:simulation_details}.
Figure ~\ref{fig:SpanningTrees} (a) shows the dependence of the success probability $P_S$  on the number of trees $M$ for $N=6$ and $p=1,2$. The results show the median of 100 different problem instances which use the same trees $T$ for decoding. The shaded area ranges from the 25th to the 75th percentile.
Increasing the number of decoding trees $M$ improves $P_S$. In addition, the error bars decrease with increasing $M$. For example, a single tree, $M=1$, reaches a median $P_S$ of about $0.5$ for $p=1$, but the error bars range from $0$ to $0.7$. This means that half of the simulated instances return a $P_S$ within this range. The outcome is unpredictable, highly depending on which tree was used for decoding. It is, therefore, advantageous to use more trees, but the improvement will saturate with increasing $M$. Further details about this are provided in Appendix~\ref{sec_app:number_of_parameters} and Fig.~\ref{fig_app:tree_scatter}. \\
Results for different system sizes $N$ with $p=1$, are shown in Fig.~\ref{fig:SpanningTrees}(b). Here, one can observe a similar behaviour as in Fig.~\ref{fig:SpanningTrees}(a), noting that with increasing system size $N$ the problem becomes harder to solve. 

\subsubsection{Logical Lines}
\label{sec:Logical_Lines}
In the previous section we studied the performance of parity QAOA using different amounts $M$ of random spanning trees. 
Now we fix the decoding to a special set of spanning trees in the parity architecture: the \textit{logical lines}~\cite{Rocchetto_2016_ScienceAdvances_LHZStabilizer,Fellner_Lechner_PhysRevLett2022_UniversalQC}, denoted by $T^{(l)}=\{t_1^{(l)}, t_2^{(l)},...t_N^{(l)}\}$. 
Logical line $i$, corresponding to tree $t_i^{(l)}$, includes all parity qubits containing the logical qubit $i$, e.g. $t_0^{(l)}$ covers the parity qubits $(0x)$ for $x=1,...,N$.
An example is $t_2$ in Fig.~\ref{fig:Decoding}.
A (all-to-all connected QUBO) problem with $N$ logical qubits has $N$ logical lines, hence $N=M$. 

\begin{figure}
\includegraphics[width=\columnwidth] {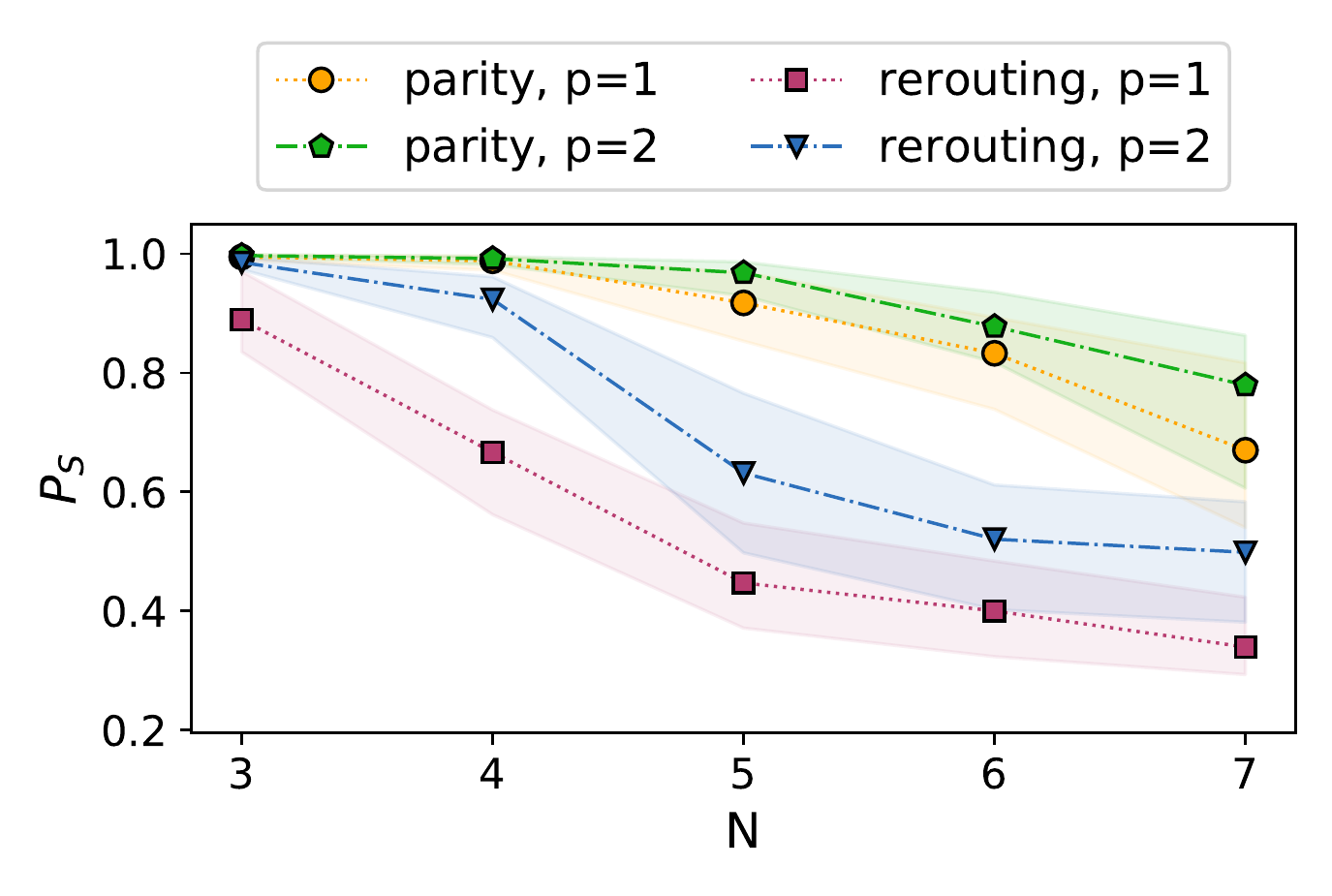}
\caption{Success probability $P_S$ in dependence of the the system size $N$, $m=N$ and  $p=1,2$. 
The dashed lines represent the median, the shaded area the 25th and 75th percentile of 100 random instances.
}
\label{fig:P_vs_n}
\end{figure}

Fig.~\ref{fig:P_vs_n} shows the success probability $P_S$ in dependence of the number of logical qubits $N$ for the different QAOA approaches.
The success probability $P_S$ for QAOA with the rerouting strategy seems to decrease slower with increasing $N$. This is due to the fact that the number of repetitions $r$ (to have the same number of qubits) increases with $N$, in fact $r=\frac{(N-1)}{2}$. $P_S$ for parity QAOA drops faster and it seems that the two curves will coincide for higher $N$. Here, we recall that $M=N$, but the total number of possible spanning trees scales exponentially with $N$. 
To keep an advantage it is necessary to increase $M$ more than linearly with $N$ for parity QAOA. 
How $P_S$ for different sizes $N$ scale with $M$ is shown in Fig. \ref{fig:SpanningTrees} (b).

\subsection{Noisy simulation}
\label{sec:noise}
In this section we investigate the performance with noisy gates, i.e. we introduce a depolarizing error on all circuit gates. The 1-qubit gate error rate will be fixed to 0.001, while the 2-qubit error will range from 0.001 to 0.1. 

With noise applied to the 2-qubit gates it is important to determine how many CNOT gates are needed in each circuit. The outlined description in the corresponding Sections \ref{sec:QAOA} and \ref{sec:LHZ_QAOA} is extended in the Appendix~\ref{app:sec:embedding}. There, it is shown that for a problem with $N=6$ qubits the rerouting layout has the advantage of fewer CNOT gates in the circuit ($94$ vs $104$ for $p=2$). 

\begin{figure}
    \includegraphics[width=\columnwidth] {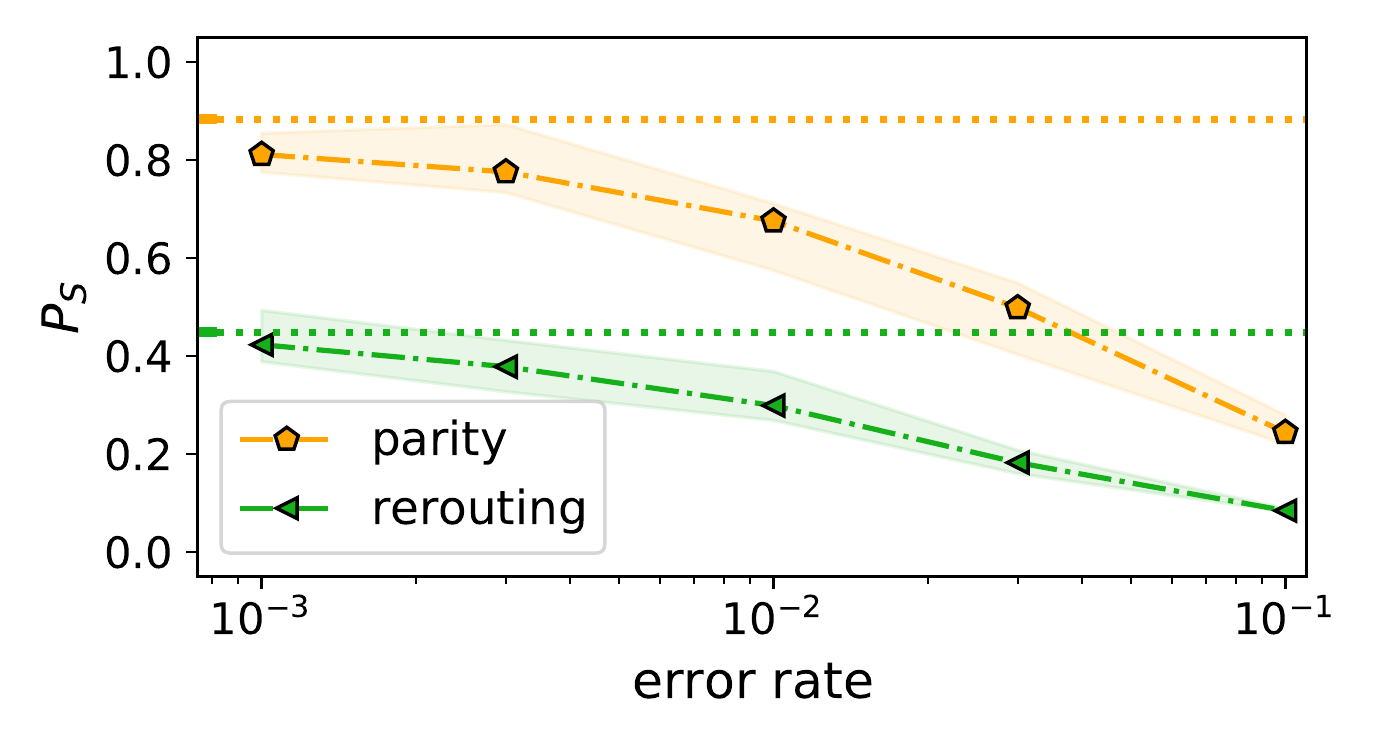}
\caption{Success probability $P_S$ for different 2-qubit gate error rates for $p=2$, $N=6$. The 1-qubit gate error rate is fixed to $0.001$. Parity QAOA is decoded by $M=N=6$ spanning trees and uses 104 CNOT gates, the circuit for QAOA with rerouting includes 94 CNOT gates. The horizontal dashed lines represent the results for an ideal circuit.
The Median of 20 random instances is shown, the shaded area is the 25th and 75th percentile.}
\label{fig:n6_Noise}
\end{figure}

Parity QAOA is decoded by the set of trees ${T^{(l)}=\{t_1^{(l)}, t_2^{(l)},...t_N^{(l)}\}}$, referred to as logical lines, as in Sec. \ref{sec:Logical_Lines}.
The obtained simulated and calculated results are shown in Fig.~\ref{fig:n6_Noise}. As seen in the previous results, parity QAOA achieves a higher success probability $P_S$ than QAOA with rerouting and is able to keep this advantage for all error rates ($<0.1$), despite having the disadvantage of a higher CNOT gate count.
With an error rate of $0.03$ the parity QAOA performs as well as the rerouting with no noise, executed $2.5$ times.

\section{Conclusions}
In this work we have shown how to apply a decoding scheme into the parity QAOA optimization routine, which exploits the redundant information introduced by the parity transformation. This novel approach leads to a better success probability $P_S$ compared to the standard rerouting method even when the number of used qubits is considered the same for both methods. For the studied problems QAOA with rerouting has the advantage of requiring fewer CNOT gates, but simulations with noise on small system sizes show that parity QAOA can keep the advantage even with high error rates ($<0.1$). 
The work in Ref.~\cite{Fellner_2021_arXiv} indicates that this advantage will disappear for bigger systems and  higher-order interaction terms, making the parity architecture more favourable. In addition, Ref.~\cite{Unger_Lechner_arXiv2022_LowDepthCircParity} shows how the reduce the gate count in parity QAOA further by some replacing CNOT gates by native ZZ gates. However, ongoing work will improve compiler quality for the rerouting circuit, hence, future work should investigate gate and constraint optimization in the parity architecture, especially for complete graphs.
Nevertheless, it is an open question how present the advantage in $P_S$ of the new introduced method will be for bigger systems sizes and how many spanning trees are the optimal choice. The results in this work suggest that the number of trees used for decoding has to be bigger than the number of logical qubits for $N>6$. There exist an exponential number of spanning trees for a complete graph, meaning that there would be enough resources. However, the classical overhead should be kept in mind. 
Although we only presented the decoding scheme for complete graphs, the extension to more general graphs with higher-order interactions is straightforward~ \cite{EnderTerHoeven_2021_arXiv_Compiler}. Those problems may have less qubit overhead, but then fewer spanning trees exist for decoding. Future work should investigate this trade-off.
In addition, it might be advantageous to adapt the objective function if one increases the system size. In this work, we used the mean energy of all used spanning trees, but a different formulation could be more beneficial considering bigger problem instances.

\section{Acknowledgements} Work was supported by the Austrian Science Fund (FWF) through a START grant under Project No. Y1067-N27 and the SFB BeyondC Project No. F7108-N38. This project was funded within the QuantERA II Programme that has received funding from the European Union’s Horizon 2020 research and innovation programme under Grant Agreement No 101017733.


\clearpage
\newpage
\appendix

\section{QAOA on planar chips 
\label{app:sec:embedding}}

In this section we describe in more detail the embedding of the different QAOA strategies on digital quantum devices. This is important when we introduce noise and the number of error-prone 2-qubit gates (CNOT) matter. \\
Here, we consider the two approaches outlined in Sec.~\ref{sec:QAOA} and \ref{sec:LHZ_QAOA}: QAOA with rerouting and parity QAOA. The new parity QAOA scheme described in Sec.~\ref{sec:log_QAOA} introduces a novel modification for the optimizer and read-out, and, therefore, uses the same embedding as the original version. 
For both methods we consider a planar chip with nearest neighbor connectivity to be in line with modern state-of-the-art quantum devices \cite{Arute_2019_Nature, Henriet_2020_Quantum, Saffman_2010_RevModPhysics, Bloch_2008_RevModPhysics}. 

\begin{figure}[h!]
  \begin{center}
  \subfloat[]{\includegraphics[]{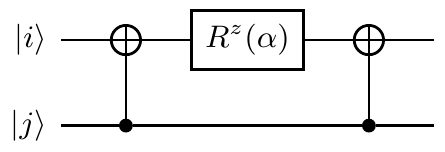}
  }
\subfloat[]{\includegraphics[]{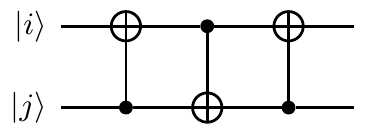}
   }
    \caption{Interactions between (neighbouring) qubits $i$ and $j$ is realized with two CNOT gates and a single qubit z-Rotation, as shown in (a). A SWAP gate, shown in (b), is implemented via 3 CNOT gates.}
    \label{fig:app:Interaction_and_SWAP}
  \end{center}
\end{figure}

The problem Hamiltonian for the rerouting strategy, Eq.~\eqref{eq:Hp-qubo}, is described by 2-body interactions. 
The corresponding phase separation unitary operator
\begin{align} 
    U_p(\gamma) &= e^{-i\gamma\hat{H}_p} = \prod_{i=1}^{N}\prod_{j<i}e^{-i\gamma J_{ij}\sigma_i^z\sigma_j^z},
\end{align}
 implements each interaction with two CNOT gates and a single $R^z(\alpha)$ rotation, as shown in Fig.~\ref{fig:app:Interaction_and_SWAP}(a). The rotation angle $\alpha$ is determined by the product of the variational parameter $\gamma$ and the interaction $J_{ij}$: ${R^z_j(\alpha)=e^{-i \alpha \sigma_j^z}}$ with $\alpha=\gamma J_{ij}$.
To be able to realize all qubit interactions non-neighbouring qubits need to be made local at least once via SWAP gates. A SWAP gate is shown in Fig.~\ref{fig:app:Interaction_and_SWAP}(b). It consists of 3 CNOT gates, which makes them resource intensive. As the problem graphs we study in this paper are fully connected, every qubit needs to interact with every other qubit.
In order to minimize the usage of SWAP gates we optimize the circuit with the t$\ket{\mathrm{ket}}$ transpiler by Cambridge Quantum Computing (CQC) \cite{tket}.

In the main text, Sec.~\ref{sec:noise}, we investigate a problem with $N=6$ qubits, results shown in Fig.~\ref{fig:n6_Noise}. 
For a circuit of this size the $t\ket{ket}$ transpiler returns $94$ CNOT gates for $p=2$, arranged on a $2\times3$ qubit lattice. (Note that the compiler did not return the most optimal circuit.)

\begin{figure}[h!]
  \begin{center}
  \includegraphics[]{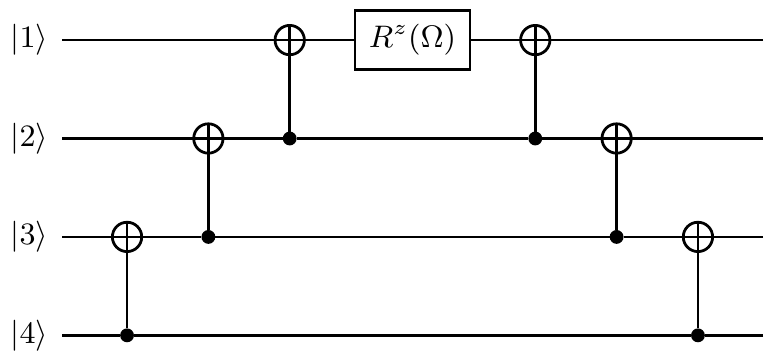}
    \caption{Circuit to implement a 4-body constraint. It consists of 6 CNOT gates and a single qubit $R^z$ rotation. In case of a 3-body constraint the last line, qubit 4, is not required.}
    \label{fig:app:Constraint_gates}
  \end{center}
\end{figure}

In the parity architecture we transformed the qubit interactions $J_{ij}$ to local fields $\Tilde{J}_{\nu}$, see Eq.~\eqref{eq:HZ}. The local field unitary,  $\tilde{U}_z(\gamma) = \prod_{\nu=1}^K e^{-i\gamma \Tilde{J}_\nu \Tilde{\sigma}_{\nu}^z}$ consists of single qubit rotations, where the rotation angle is determined by $\gamma$ and $\Tilde{J}_{\nu}$.
However, the constraint unitary, 
\begin{align}  \label{eq:HP_log}
 \tilde{U}_c(\Omega) &= \prod_{l=1}^{L} e^{-i\Omega H_{\squaredots,l}}=\prod_{l=1}^{L}e^{-i\Omega \Tilde{\sigma}_{(l,1)}^z\Tilde{\sigma}_{(l,2)}^z\Tilde{\sigma}_{(l,3)}^z[\Tilde{\sigma}_{(l,4)}^z]},
\end{align}
needs to be realized via CNOT gates. The unitary contains 3 or 4 $\sigma^z$ terms, depending whether it is a 3- or 4-body constraint.
4-body (3-body) constraints consist of a consecutive sequence of 3 (2) CNOT gates, a $R^z(\Omega)$ rotation and 3 (2) CNOT gates back on the same qubits. That are in total 6 (4) CNOT gates per 4-body (3-body) constraint. This sequence is visualized in Fig.~\ref{fig:app:Constraint_gates}. 

The problem studied in the main text, Sec.~\ref{sec:noise}, results shown in Fig.~\ref{fig:n6_Noise}, has $K=15$ qubits in the parity representation. In total it has six 4-body and four 3-body constraints. This sums up to 52 CNOT gates for for each layer, resulting in $104$ CNOT gates for $p=2$. 

In this example QAOA, with rerouting has the advantage of requiring fewer CNOT gates ($94$ vs $104$ CNOT gates).
However, Ref. \cite{Fellner_2021_arXiv} gives a more detailed comparison about the two mention models and used CNOT gates for different optimization problems. There, the result suggest a significant advantage of the parity encoding for bigger system sizes and higher order interaction terms. In this work we study small system sizes and two-qubit interactions only. 
The described implementation of constraints in parity QAOA can be further enhanced, as shown in the recent work of Ref.~\cite{Unger_Lechner_arXiv2022_LowDepthCircParity}. This improvement was not considered in this work or in Ref.~\cite{Fellner_2021_arXiv}.

\section{Decoded parity QAOA}
\label{sec_app:dec_parity}
\begin{figure}
\includegraphics[scale=0.52]{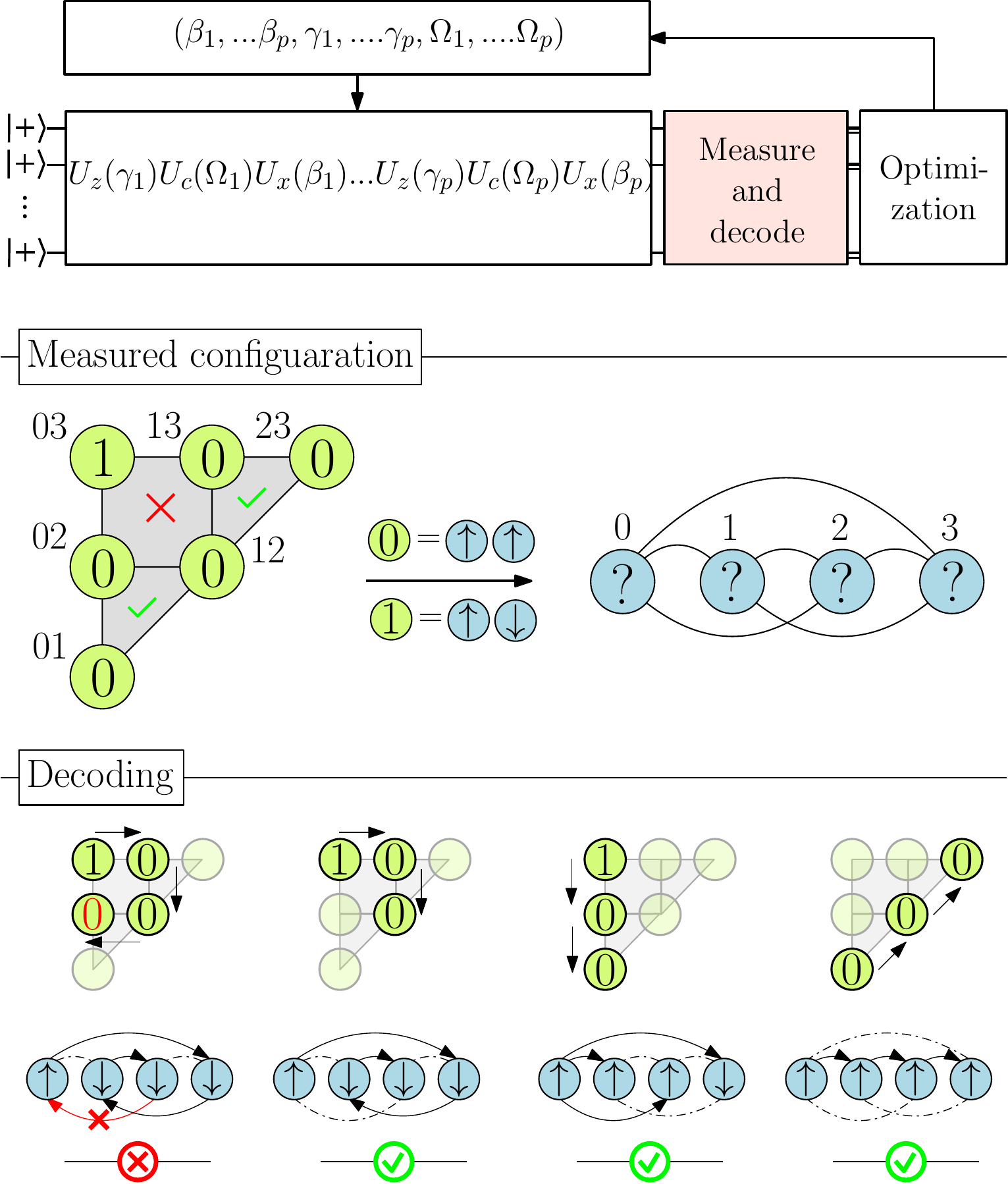}
\caption{QAOA on the parity architecture with decoding. On top, the circuit applies the unitarys as usual, with $U_z(\gamma)$ and $U_c(\Omega)$ as problem terms and $U_x(\beta)$ as the driver term. With the measurement, physical configurations are decoded into logical ones, such that a logical energy serves as objective function to the classical optimizer. 
An example of a physical configuration (green qubits) is given in the middle. The configuration violates a constraint (red cross) and therfore has no corresponding logical state (blue qubits). 
4 decoding examples are shown in the bottom. In the first example the 4 qubits of the 4-body constraint are used for decoding. With the last physical qubit one runs into a contradiction, not all parities of the logical qubits can be fulfilled. The examples shown take physical qubits for decoding that relate to a spanning the in the logical one, marked as black arrows in the logical graph. With this subset of physical qubits it is possible to construct a valid logical state.}
\label{fig_app:Decoding}
\end{figure}

This chapter gives a more detailed description of the decoding strategy for parity QAOA (Sec.\ref{sec:log_QAOA}).
The adapted QAOA procedure is shown on top of Fig.~\ref{fig_app:Decoding}.
The unitaries are applied on an initial state  $\ket{\tilde{\Psi}_0}=\ket{+}^{K}$ as described in Eq.~\eqref{eq:Parity_circuit}. This is followed by a measure and decoding scheme, which returns a logical, and not a physical energy to the optimizer. The optimizer then feeds the adapted parameters back into the circuit.

To obtain a logical energy of a physical configuration it is necessary to decode the physical configuration $\tilde{\bfq}$ into a logical one $\bfq$. As mentioned in Sec.~\ref{sec:LHZ_QAOA}, configurations in the parity architecture which violate the constraints, have no corresponding logical configuration. Such an unphysical state is shown as the measured configuration in the middle of Fig.~\ref{fig:Decoding} (green qubits). This configuration fulfils the 3-body constraints (green check), but not the 4-body constraint (red cross). Therefore, we can not construct a logical configuration $\bfq$ with the usual decoding rules: $0\to \;\uparrow \uparrow$ or $\downarrow \downarrow$ (parallel logical qubits), $1 \to \;\uparrow \downarrow$ or $\downarrow \uparrow$ (anti-parallel logical qubits).
To see this, consider the first decoding example on the bottom left of Fig.~\ref{fig:Decoding}: The decoding starts at the top left qubit $03$. It has the value 1 and therefore the logical qubits $0$ and $3$ have to be anti-parallel. We can set $0$ to up and $3$ to down. The next two physical qubits, $13$ and $12$, give the information that qubit $1$ is parallel to qubit $3$ and qubit $2$ is parallel to $1$. We can set the qubits $1$ and $2$ to down. The fourth and last physical qubit, $02$, has the value $0$, this means qubit $0$ and $2$ should be parallel, but they are already anti-parallel. There is no valid logical configuration that fulfills all parity qubits. 
Before considering the last physical qubits ($02$) all the logical qubits were determined. This is the second example in the figure. Here, 3 of the 4 previous qubits are taken for the decoding. The 3 physical qubits correspond to the interactions, marked as arrows, in the logical graph. Those interactions cover all the qubits without making a cycle (The previous example made a cycle). This is the definition of a spanning tree in a graph. The last two examples show different spanning trees and their corresponding decoded state.
Note that they do not return the same logical state as the corresponding physical configuration violates a constraint.

\section{Simulation details}
\label{sec_app:simulation_details}
The QAOA simulation details are the following:
To optimize the QAOA variational parameters $\bfbeta$ and $\bfgamma$ and $\bfbeta, \bfgamma$ and $\bfOmega$ for rerouting QAOA and parity QAOA respectively, we initialize 100 random start parameters (as proposed in Ref.~\cite{Shaydulin_2019}) followed by parameter updates. Updates are accepted with the Metropolis criterion at a constant temperature.
In general, there will be 10.000 measurements for each step. We simulate 100 problem instances for the noiseless and 20 instances for the noisy simulations, whereas the best run (lowest energy $\calC(\bfbeta^*,\bfgamma^*)$ or $\Tilde{\calC}(\bfbeta^*,\bfgamma^*,\bfOmega^*)$) out of the 100 random initialization is taken for latter statistics.
For the simulations with noise the number of measurements during the optimization is set to 1,000 and readjusted to 10,000 for executing the circuit with the best found parameters $\bfbeta^*$ and $\bfgamma^*$ or $\bfbeta^*, \bfgamma^*$ and $\bfOmega^*$.
With the choice $J_{ij}\in\{\pm 0.1, \pm0.2,...,\pm1.0\}$ the search space of parameter $\gamma$ is restricted to $[0, 10\pi)$. The search space of the other parameters is limited to $[0, \pi)$. 

The set of random spanning trees $T = \{t_1,t_2,\dots\,t_M\}$ for decoded parity QAOA are generated the following way: One starts with a random edge, containing two nodes. One of these nodes is randomly chosen to continue the tree. The next new edge is chosen randomly from the set of possible edges, excluding cycles. This procedure is repeated until all nodes are included.

\section{Comparing different strategies}
\label{sec_app:different_strategies}
In this section we use a different measure for the success probability $P_S$ than outlined in Sec.~\ref{sec:log_QAOA}. 

In the main text, after measuring the optimal variational state $\ket{ \tilde{\Psi}(\bfbeta^*,\bfgamma^*,\bfOmega^*)}$ we returned the best decoded state  $\boldsymbol{q}^{T}$ defined in Eq.~\eqref{eq:outputted_decoded_string}.
The argument is that at the measurement in the very end one can keep the decoded state with the lowest energy and discard the others. 
However, in this section, we want to keep all the decoded states and consider the probability to measure the ground state within them. In this case Eq.~\eqref{eq:outputted_decoded_string}. Instead, the output distribution is described by the quantum state $\ket{D_{T}\tilde{\Psi}(\bfbeta^*,\bfgamma^*,\bfOmega^*)}$, where $D_{T}$ is the linear decoding map $\ket{D_{T}\tilde{\bfq}} = \frac{1}{\sqrt{M}} \big( \ket{\bfq^{(t_1)}} + \ket{\bfq^{(t_2)}} + ... + \ket{\bfq^{(t_M)}}\big)$. The definitions of the ground state probability Eq.~\eqref{eq:ground_state_probability} and the success probability Eq.~\eqref{eq:mod_PGS} do not change. 
We compare parity QAOA as described in the main text with the here defined `mean parity' QAOA method. In both cases we set $r=1$ in Eq.~\eqref{eq:mod_PGS}. 

In addition, we also look at the rerouting method, setting the number of repetitions $r=M$ of Eq.~\eqref{eq:mod_PGS}, compared to $r=N/K$ in the main text.

\begin{figure}
\includegraphics[width=\columnwidth]{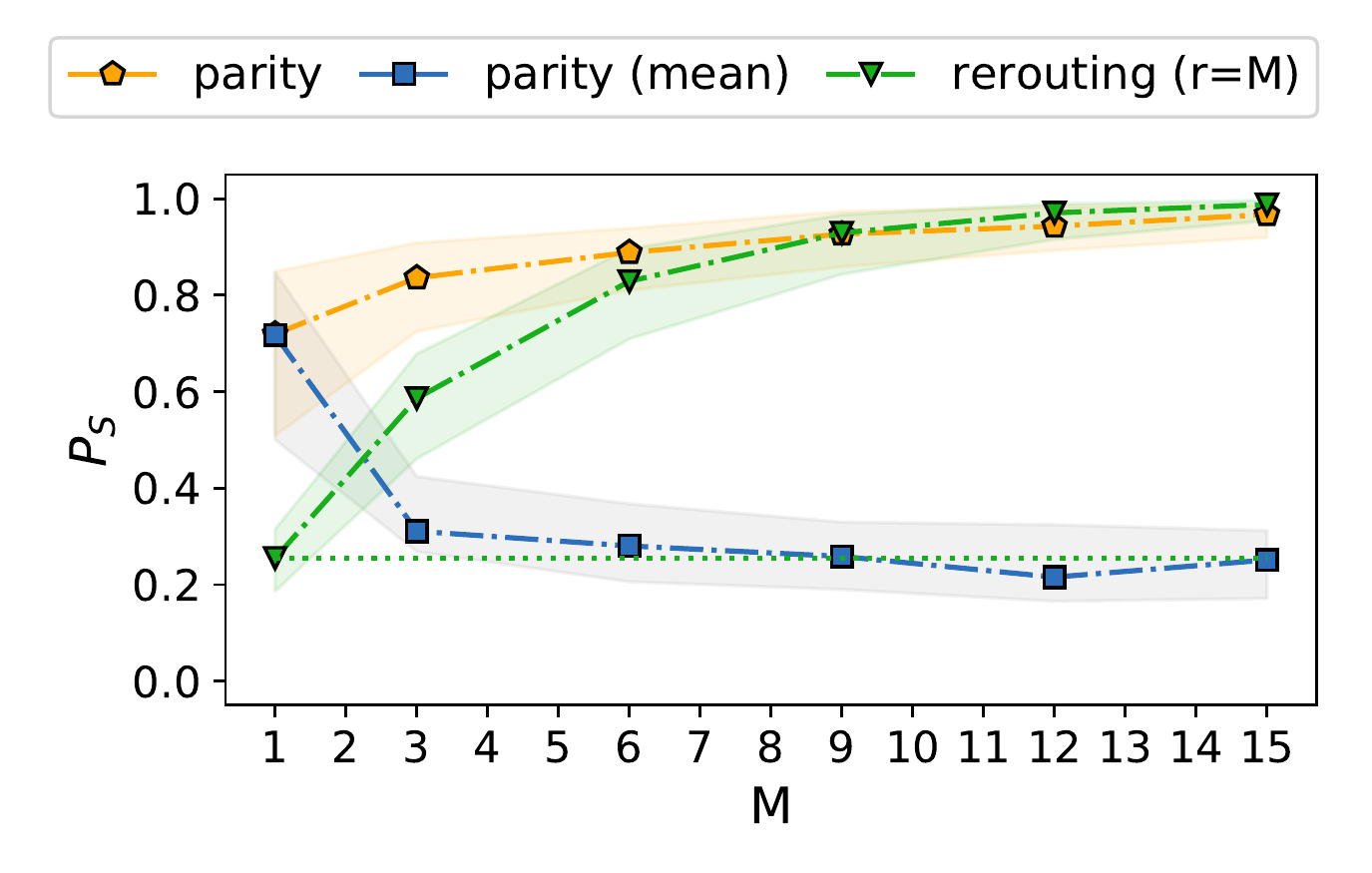}
\caption{Success probability $P_S$ in dependence of the number of spanning trees $M$ used, $N=6$, $p=2$.
When creating the variational state $\ket{ \tilde{\Psi}(\bfbeta^*,\bfgamma^*,\bfOmega^*)}$ with the best found parameters, Parity QAOA keeps the best decoded logical state per physical state while w-parity QAOA keeps all decoded logical states to determine the success probability $P_S$.
For the rerouting strategy the results for $P_S$ are adapted with Eq.~\eqref{eq:mod_PGS} with $r=M$.}
\label{fig_app:SpTree}
\end{figure}

In Fig.~\ref{fig_app:SpTree} we can see the results for the success probability $P_S$ for parity, mean parity and rerouting, for $N=6$. 
For $M=1$ both parity methods return the same $P_S$. For increasing $M$ the probability for mean parity drops and converges to a value of $P_S\sim0.3$.
Similar $P_S$ is achieved by the rerouting method, with a single copy (green dashed line).
 
According to the results, the rerouting approach needs to be repeated at least $M=9$ times to reach the same or more $P_S$ than parity QAOA.

\section{Parity QAOA+}
\label{sec:modular}
\begin{figure}
\includegraphics[width=\columnwidth]{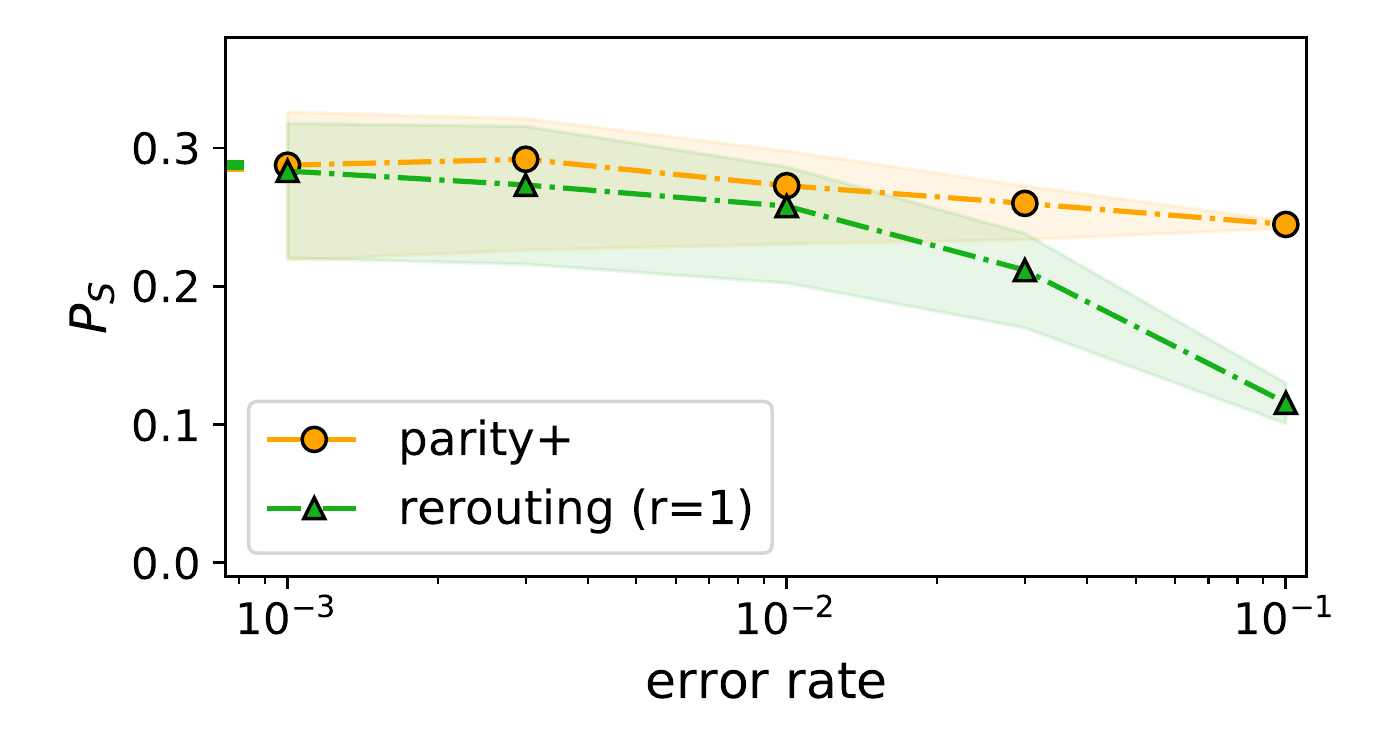}
\caption{Noise on parity QAOA+ and QAOA with rerouting, $N=5$, $p=1$. The 1-qubit gate error rate is fixed to $0.001$.}
\label{fig_app:Modular}
\end{figure}

There are several adaptions to the simple p-level QAOA method described in the main text in Sec.~\ref{sec:QAOA} \cite{Hadfield_2019_Algorithms_QAOA+, Bako_Zimboras_arXiv2022_FUNCQAOA, Wang_Rieffel_PhysRevA2020_XYMixersQAOAPlus, Herrman_SciReports2022_MultiAngleQAOA,Yu_arXiv2022_QAOAwithAdaptiveBiasFields, Bravyi_PhysRevLett2020_RQAOAMaxCut,Bravyi_arXiv2022_RQAOAColor,Egger_Quantum2021_warmstartingquantum,Govia_Krovi_AmericanPhysSoc_2021_MixerRotAxis,Zhu_Economou_PhysRevResearch2022_AdaptQAOA}. For instance, the Quantum Alternation Operator Ansatz~\cite{Hadfield_2019_Algorithms_QAOA+} (QAOA+) provides a more expressive variational state by modifying the definitions of $U_x(\beta)$ and $U_p(\gamma)$.
In Ref.~\cite{Ender_2022_PRX_ModularParity} the authors propose a novel approach for parity QAOA, which is based on QAOA+. Here, one starts with constraint fulfilling states and the driver Hamiltonian $\tilde{H}_x^{(\mathrm{mod})}$ is adapted such that it only allows transitions between constraint fulfilling states (in an ideal circuit). No parity constraints required. 
This gives the advantage of increased performance, but with the drawback of non parallelizable gates. With an hybrid approach, where some parity constraints are enforced implicit with the driver Hamiltonian and the others are applied explicit with parity constraints, one obtains a trade-off between performance and parallelizability of the gates.
Nevertheless, here we want to focus on the first. We use a driver unitary $\tilde{U}_x^{(\mathrm{mod})}(\beta) =  e^{-\beta\tilde{H}_x^{(mod)}}$ such that the states during the parity QAOA procedure stay in the constraint fulfilling subspace.
As in the main text, we study QUBO problems with all-to-all connectivity, here with $N=5$.
In Sec.~\ref{sec:Logical_Lines} we defined a set of spanning trees that correspond to the logical lines in the parity architecture, namely $T^{(l)}=\{t_1^{(l)}, t_2^{(l)},...t_N^{(l)}\}$ with $t_i^{(l)}$, including all parity qubits containing the logical qubit $i$.
Now one can define a driver term with
\begin{equation}
    \label{eq:driver_line}
    X_i = \prod_{k \in t_i^{(l)}}\sigma_x^{(k)}.
\end{equation}
The new driver Hamiltonian is defined as the sum over the driver terms $\tilde{H}_x^{\mathrm{mod}} = \sum_{i=1}^{N} X_i$.
With this Hamiltonian, parity QAOA+ and QAOA with rerouting are mathematically equivalent. 
Here, we consider single copies in both cases, hence $r=1$ in Eq.~\eqref{eq:mod_PGS}. Simulations on an ideal circuit yield the same output. This can be seen in Fig.~\ref{fig_app:Modular}, where the results for the success probability $P_S$ for $N=5$ and $p=1$ are shown. The solid bars on the right represent the results with no noise. As soon as a depolarizing 2-qubit gate error is applied, the simulation results between the different approaches differ. Here, parity+ needs more CNOT gates than the rerouting layout. Nevertheless, applying the decoding to parity+ shows better noise stability.

\section{Number of parameters and spanning trees}
\label{sec_app:number_of_parameters}
\begin{figure}
\includegraphics[width=\columnwidth]{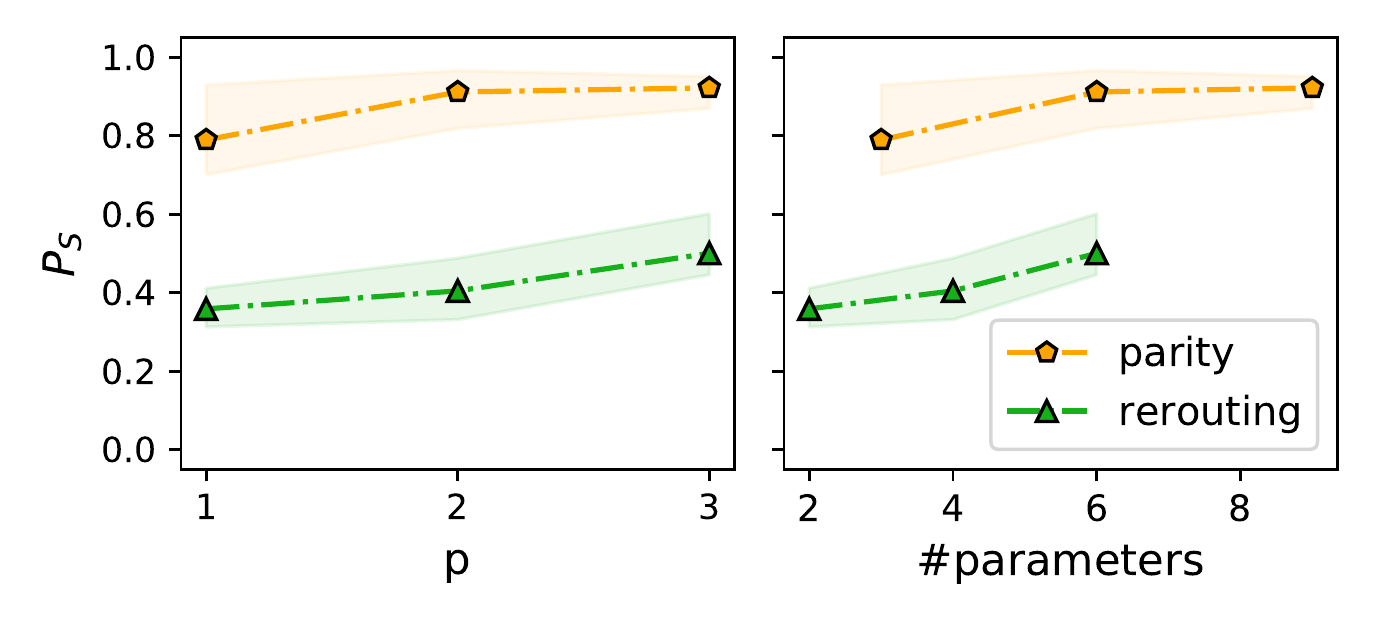}
\caption{Success probability $P_S$ in dependence of the circuit depth p (left) and in dependence of the total number of parameters (right). The rerouting method needs to optimize 2 parameters per cycle ($\bfbeta$ and $\bfgamma$), the parity method 3 ($\bfbeta,\bfgamma$ and $\bfOmega$). The number of parameter updates is $100/500/1100$ for $p=1/2/3$. The number of measurements is set to $1000$ during optimization and to $10.000$ for the final measurement.}
\label{fig_app:PS_vs_number_of_paraemters}
\end{figure}

Figure~\ref{fig_app:PS_vs_number_of_paraemters} (left) shows the success probability in dependence of the circuit layers $p$. Nevertheless, one needs to keep in mind that the different QAOA approaches have different numbers of parameters associated to each depth.
The variational state created by the rerouting method, Eq.~\eqref{eq:QAOA_circuit}, depends on $2p$ parameters $\bfbeta$ and $\bfgamma$. In contrast, the parity circuit creates the variational state, Eq.~\eqref{eq:Parity_circuit}, with $3p$ parameters $\bfbeta,\bfgamma$ and $\bfOmega$. This means that the rerouting strategy has less parameters to optimize, e.g., one can perform QAOA with depth $p=3$ and one has as many parameters to optimize as with parity QAOA with depth $p=2$. This is shown in Fig.~\ref{fig_app:PS_vs_number_of_paraemters} (right). We can see that the rerouting method with $p=3$ (and 1100 parameter updates) still cannot reach the same success probability as the parity method with $p=2$ (500 parameter updates).

\begin{figure}
\includegraphics[width=\columnwidth]{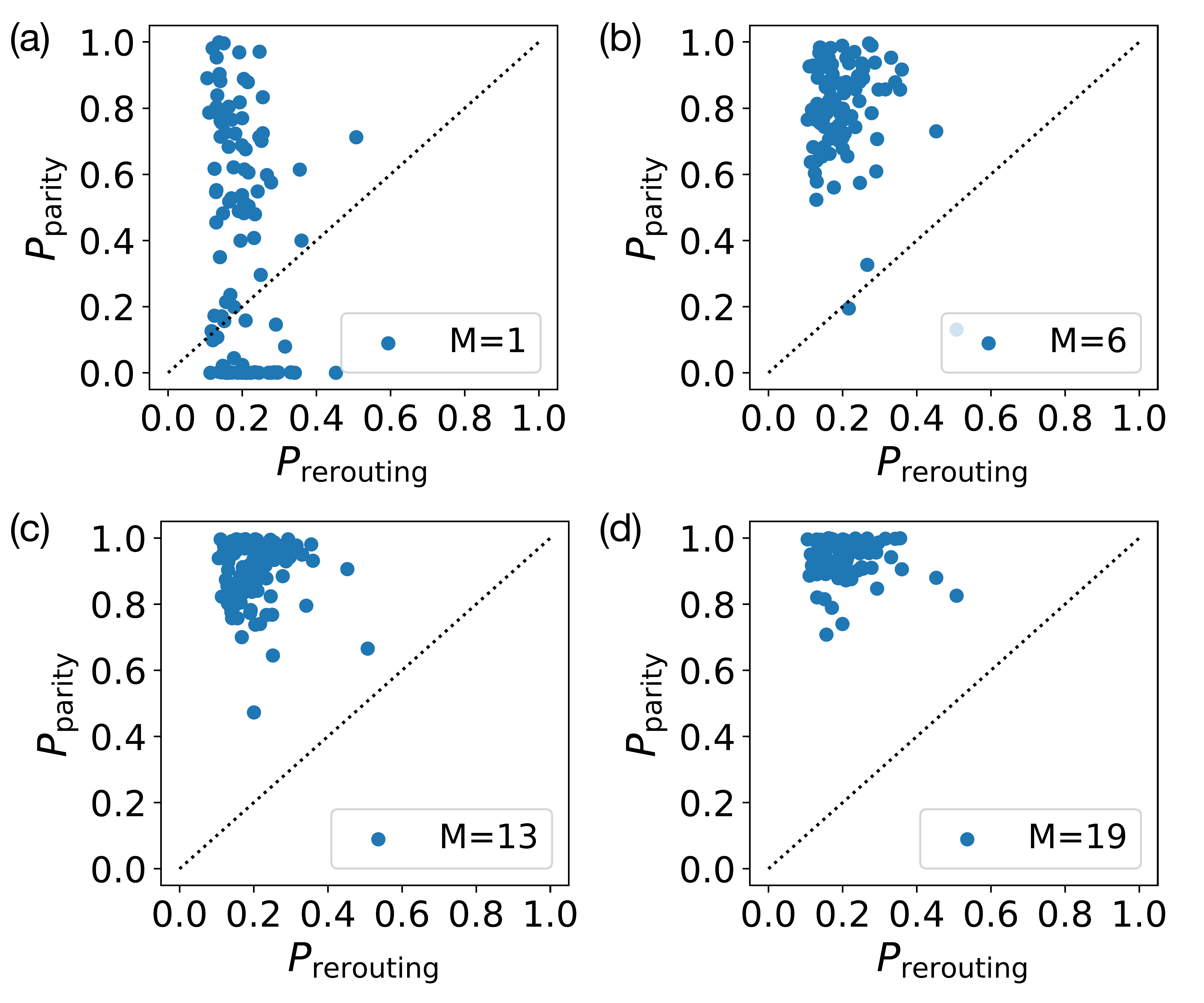}
\caption{Success Probabilities of the rerouting ($P_{\mathrm{rerouting}}$) and parity strategy ($P_{\mathrm{parity}}$) for different amount $M$ of random spanning trees, $N=6$, $p=1$. One tree for decoding, ${M=1}$ returns unpredictable success rates for parity QAOA, as shown in (a), while QAOA with rerouting gives a stable performance. With increasing $M$, $M=6$ in (b), $M=13$ in (c) and $M=19$ in (d), this problem for parity QAOA will disappear.}
\label{fig_app:tree_scatter}
\end{figure}

In Sec.~\ref{sec:Spanning_trees}, we looked at the performance of parity QAOA in dependence of the number of spanning trees $M$ used for decoding. Here we want to look at the 100 instances individually and, in addition, compare it to the rerouting method, which is independent of $M$. Figure~\ref{fig_app:tree_scatter} shows the success probabilities of the two different methods for different $M$. In panel (a) we see the results for $M=1$. The results for parity QAOA ($P_{\mathrm{parity}}$) are distributed over the whole interval $[0,1]$, meaning that the outcome is unpredictable. On the other hand, QAOA with rerouting returns stable results. This disadvantage for parity QAOA will disappear with increasing $M$, meaning that it will be necessary to include enough spanning trees in the decoding. Figure~\ref{fig_app:tree_scatter} shows the outcome of $M=6, 13, 19$ in panels (b), (c) and (d).
%

\end{document}